\documentclass[onecolumn]{emulateapj}  

\shorttitle{Compact Rotating Dense Molecular Gas Torus in NGC 1068}    
\shortauthors{Imanishi et al.}

\begin{document}


\title{ALMA Reveals an Inhomogeneous Compact Rotating Dense
Molecular Torus at the NGC 1068 Nucleus} 

\author{Masatoshi Imanishi\altaffilmark{1}}
\affil{National Astronomical Observatory of Japan, National Institutes 
of Natural Sciences (NINS), 2-21-1 Osawa, Mitaka, Tokyo 181-8588, Japan} 
\email{masa.imanishi@nao.ac.jp}

\author{Kouichiro Nakanishi\altaffilmark{1}}
\affil{National Astronomical Observatory of Japan, National Institutes 
of Natural Sciences (NINS), 2-21-1 Osawa, Mitaka, Tokyo 181-8588, Japan}

\author{Takuma Izumi}
\affil{National Astronomical Observatory of Japan, National Institutes 
of Natural Sciences (NINS), 2-21-1 Osawa, Mitaka, Tokyo 181-8588, Japan}

\and 

\author{Keiichi Wada} 
\affil{Kagoshima University, Kagoshima 890-0065, Japan}

\altaffiltext{1}{Department of Astronomical Science, SOKENDAI (The
Graduate University of Advanced Studies), 2-21-1 Osawa, Mitaka, Tokyo
181-8588, Japan}

\begin{abstract}
We present the results of our ALMA Cycle 4 high-spatial-resolution
(0$\farcs$04--0$\farcs$07) observations, at HCN J=3--2 and HCO$^{+}$
J=3--2 lines, of the nucleus of NGC 1068, the nearby prototypical type 2 
active galactic nucleus (AGN).
Our previous ALMA observations identified the compact emission of
these lines at the putative location of the torus around a
mass-accreting supermassive black hole.
We now report that we have detected the rotation of this compact
emission, with the eastern and western sides being redshifted and
blueshifted, respectively. 
Unlike the previously reported CO J=6--5 emission, both the
morphological and dynamical alignments of the HCN J=3--2 and HCO$^{+}$
J=3--2 emission are roughly aligned along the east--west direction
(i.e., the expected torus direction), suggesting that these molecular
lines are better probes of a rotating dense molecular gas component
in the torus. 
The western part of the torus exhibits larger velocity dispersion and
stronger emission in the HCN J=3--2 and HCO$^{+}$ J=3--2 lines than
the eastern part, revealing a highly inhomogeneous molecular torus. 
The dense molecular gas in the torus and that of the host galaxy at 
0$\farcs$5--2$\farcs$0 from the AGN along the torus direction are found
to be counter-rotating, suggesting an external process happened in the
past at the NGC 1068 nucleus.     
\end{abstract}


\section{Introduction}

Active galactic nuclei (AGNs) exhibit very
bright emission from the nuclear compact regions of galaxies.
The origin of the emission is believed to be the release of
gravitational energy by a mass-accreting supermassive black hole (SMBH) 
into radiative energy. 
Some AGNs exhibit a strong time-varying optical continuum and broad
optical emission lines (classified as type 1), while others do not
(type 2). 
A unified picture of AGNs has been proposed based on observations of the
nearby type-2 AGN, NGC 1068 (z = 0.0037, distance $\sim$ 14 Mpc, 1
arcsec is $\sim$70 pc) \citep{ant85}; Type-1 and -2 AGNs are
intrinsically the same, but the central engine (= a mass-accreting SMBH)
and the surrounding sub-pc-scale broad-line-emitting regions of a type-2
AGN are hidden behind a toroidally distributed obscuring medium, the
so-called ``dusty molecular torus'' \citep{ant93}. 
This AGN unification paradigm has become widely accepted because it
naturally explains observational results from various AGNs.  
In this way, the torus plays an important role in AGNs, but our
understanding of its 
physical/chemical/morphological/dynamical properties is
observationally still highly incomplete because its compact size ($<$a
few 10 pc or $<$0.5 arcsec at a $>$10 Mpc distance) makes it difficult
to clearly spatially resolve emission from the torus. 

The dust in close proximity to the mass-accreting SMBH is heated
to high temperatures ($>$100 K) by the AGN's strong energetic UV--X-ray 
radiation, so that the torus is expected to shine brightly in the
mid-infrared (3--20 $\mu$m) wavelengths.  
Using the very high-spatial-resolution ($\sim$10 mas) interferometric 
technique, the mid-infrared $\sim$10 $\mu$m emitting region of the torus
in NGC 1068 was estimated to be $<$several pc elongated along the 
east--west direction \citep{jaf04,pon06,rab09,bur13,lop14}.    
However, the $\sim$10 $\mu$m continuum emission may be biased toward the
inner hotter region and therefore may not reflect the overall structure
of the torus \citep{sch08}.  
Furthermore, recent mid-infrared observations of NGC 1068 and other
nearby AGNs have demonstrated that the observed mid-infrared continuum
is often dominated by radiation from and/or scattering by polar dust,
rather than emission from the putative torus
\citep{boc00,alo00,tom01,tri14,hon17}.
Quantitative discussion of the torus based on mid-infrared continuum
observations has recently been recognized as more complicated than
previously thought. 
Finally, and most importantly, no dynamical information can be obtained
from the dust continuum data.

High-spatial-resolution molecular line observations using ALMA are a
powerful tool for investigating the morphological structure of the torus,
where molecular gas and dust are expected to coexist. 
More importantly, we can use molecular line observations to obtain
information about the dynamics of the torus.  
\citet{ima16} presented 0$\farcs$1--0$\farcs$2 resolution ALMA
observational results of the NGC 1068 nucleus at the HCN J=3--2 and 
HCO$^{+}$ J=3--2 lines (= dense molecular tracers) and detected
compact emission at the location of the putative torus
around the central mass-accreting SMBH by clearly isolating from the
surrounding much brighter molecular emission at the inner $<$5 arcsec
region of the host galaxy.  
However, no meaningful information concerning the dynamics of the
torus was obtained, due to the limited spatial resolution.
\citet{gar16} conducted 0$\farcs$05--0$\farcs$07 resolution CO J=6--5
line observations of the NGC 1068 nucleus and spatially resolved the
molecular emission from the torus.
These authors found that the redshifted and blueshifted CO J=6--5
emission lines were aligned along the north--south direction, which is
highly tilted with respect to the morphologically elongated east--west
direction (i.e., the expected torus direction), and interpreted that the
molecular gas in the torus is highly turbulent \citep{gar16}. 
\citet{gal16} also detected redshifted and blueshifted CO J=6--5 line 
emission roughly along the north--south direction and argued that the
detected emission originated from bipolar outflow activity in the
direction almost perpendicular to the torus. 
The dynamical origin of the compact CO J=6--5 line emission is not
clear, and the CO J=6--5 emission could be largely affected by highly
excited molecular gas in the outflow. 
To better understand the properties of the torus, it is desirable to
make high-spatial-resolution observations using different dense gas
tracers.  
In this letter, we present the results of our ALMA
0$\farcs$04--0$\farcs$07 resolution observations of the HCN J=3--2 and
HCO$^{+}$ J=3--2 lines of the NGC 1068 nucleus. 

\section{Observations and Data Analysis}

We conducted ALMA band 6 (211--275 GHz) observations of the NGC 1068 
nucleus in our Cycle 4 program 2016.1.00052.S (PI = M. Imanishi).
In addition to the long baseline observations required to achieve very
high-spatial-resolution (0$\farcs$04--0$\farcs$07), we conducted shorter
baseline observations to recover $\sim$1$''$ scale spatially extended 
diffuse emission.  
Table 1 summarizes our observational details.
We combined our Cycle 4 data with our ALMA Cycle 2
0$\farcs$1--0$\farcs$2 resolution observational data (2013.1.00188.S) of
the same molecular lines to improve the signal to noise (S/N) ratio. 

\begin{deluxetable}{llccc|ccc}
\tabletypesize{\scriptsize}
\tablecaption{Log of ALMA Cycle 4 and 2 Observations \label{tbl-1}} 
\tablewidth{0pt}
\tablehead{
\colhead{Data} & \colhead{Date} & \colhead{Antenna} & 
\colhead{Baseline} & \colhead{Integration} & \multicolumn{3}{c}{Calibrator} \\ 
\colhead{} & \colhead{[UT]} & \colhead{Number} & \colhead{[m]} &
\colhead{[min]} & \colhead{Bandpass} & \colhead{Flux} & \colhead{Phase}  \\
\colhead{(1)} & \colhead{(2)} & \colhead{(3)} & \colhead{(4)} &
\colhead{(5)} & \colhead{(6)} & \colhead{(7)}  & \colhead{(8)} 
}
\startdata 
Cycle 4 & 2016 Oct 19 & 44 & 17--1808 & 44 & J0238$+$1636 &
J0238$+$1636 & J0239$-$0234 \\  
Cycle 4 & 2016 Oct 19 & 44 & 17--1808 & 44 & J0238$+$1636 &
J0238$+$1636 & J0239$-$0234  \\  
Cycle 4 (H1) & 2017 Sep 11 & 40 & 41--7552 & 39 & J0006$-$0623 &
J0006$-$0623 & J0239$-$0234 \\
Cycle 2 & 2015 Sep 19 & 36 & 41--2270 & 25 & J0224$+$0659 & J0334$-$401 & J0219$+$0120 
\enddata

\tablecomments{ 
Col.(1): Data. 
Col.(2): Observing date in UT. 
Col.(3): Number of antennas used for observations. 
Col.(4): Baseline length in meter. Minimum and maximum baseline lengths are 
shown.  
Col.(5): Net on source integration time in min.
Cols.(6), (7), and (8): Bandpass, flux, and phase calibrator for the 
target source, respectively.
}

\end{deluxetable}

We adopted the widest 1.875 GHz width mode in each spectral window. 
We took spectra at 263.7--268.6 GHz with three spectral windows to 
cover HCN J=3--2 ($\nu_{\rm rest}$ = 265.886 GHz) and HCO$^{+}$ J=3--2
($\nu_{\rm rest}$ = 267.558 GHz) lines, as well as vibrationally excited 
HCN v$_{2}$=1f J=3--2 ($\nu_{\rm rest}$ = 267.199 GHz) and HCO$^{+}$
v$_{2}$=1f J=3--2 ($\nu_{\rm rest}$ = 268.689 GHz) lines.

We used CASA (https://casa.nrao.edu) for our data
reduction, starting from calibrated data provided by ALMA.
We determined a continuum flux level using data that were not affected
by obvious emission lines, and used the CASA task ``uvcontsub'' to
subtract a constant continuum from molecular line data.
We then applied the task ``clean'' (Briggs-weighting 
\footnote{
www.aoc.nrao.edu/dissertations/dbriggs/
}, 
robust=0.5, gain $=$ 0.1)
to the continuum-subtracted molecular line data and the extracted
continuum data. 
The adopted velocity resolution was $\sim$20 km s$^{-1}$, and the 
pixel scale was 0$\farcs$01 pixel$^{-1}$.

We created cleaned images by using (a) Cycle 4 longest baseline data
only (dubbed as ``H1'' in Table 1) and (b) Cycle 2 and all Cycle 4 data.
We achieved a smaller spatial resolution for (a), but recovered a larger
amount of molecular emission from the central several arcsec region of 
the host galaxy from (b).   
We thus used the data from (a) to investigate the morphological and
dynamical properties of the compact torus in detail, as the achieved 
S/N ratios were sufficient for this purpose. 
The data from (b) will be used to discuss molecular emission in the host  
galaxy outside the compact torus-associated emission, by recovering
$\sim$1$\farcs$3 scale emission components. 
Diffuse emission with a spatial extent of $>$2$''$ is resolved out 
in our ALMA data.

\section{Results}

Figure 1 shows the continuum map, integrated intensity (moment 0) and 
intensity-weighted mean velocity (moment 1) maps of HCN J=3--2 and
HCO$^{+}$ J=3--2, obtained using dataset (b).  
Their overall spatial distributions within the central $\sim$5 arcsec of
NGC 1068, previously revealed by our 0$\farcs$1--0$\farcs$2 resolution
Cycle 2 data \citep{ima16}, are reproduced in our new Cycle 4 data. 
The continuum peak position in our 0$\farcs$01 pixel$^{-1}$ map is 
(02$^{h}$42$^{m}$40.710$^{s}$, $-$00$^{\circ}$00$'$47.94$''$)ICRS for
both dataset (a) and (b). 
We regard this coordinate as the location of the AGN (i.e., a mass-accreting
SMBH) because it spatially coincides with the radio VLBA 5 GHz
continuum peak position \citep{gal04}.   
The continuum emission at $\sim$266 GHz is detected also at the northern side
of the nucleus, along the radio jet \citep{gal04}. 

\begin{figure}
\begin{center}
\includegraphics[width=7cm]{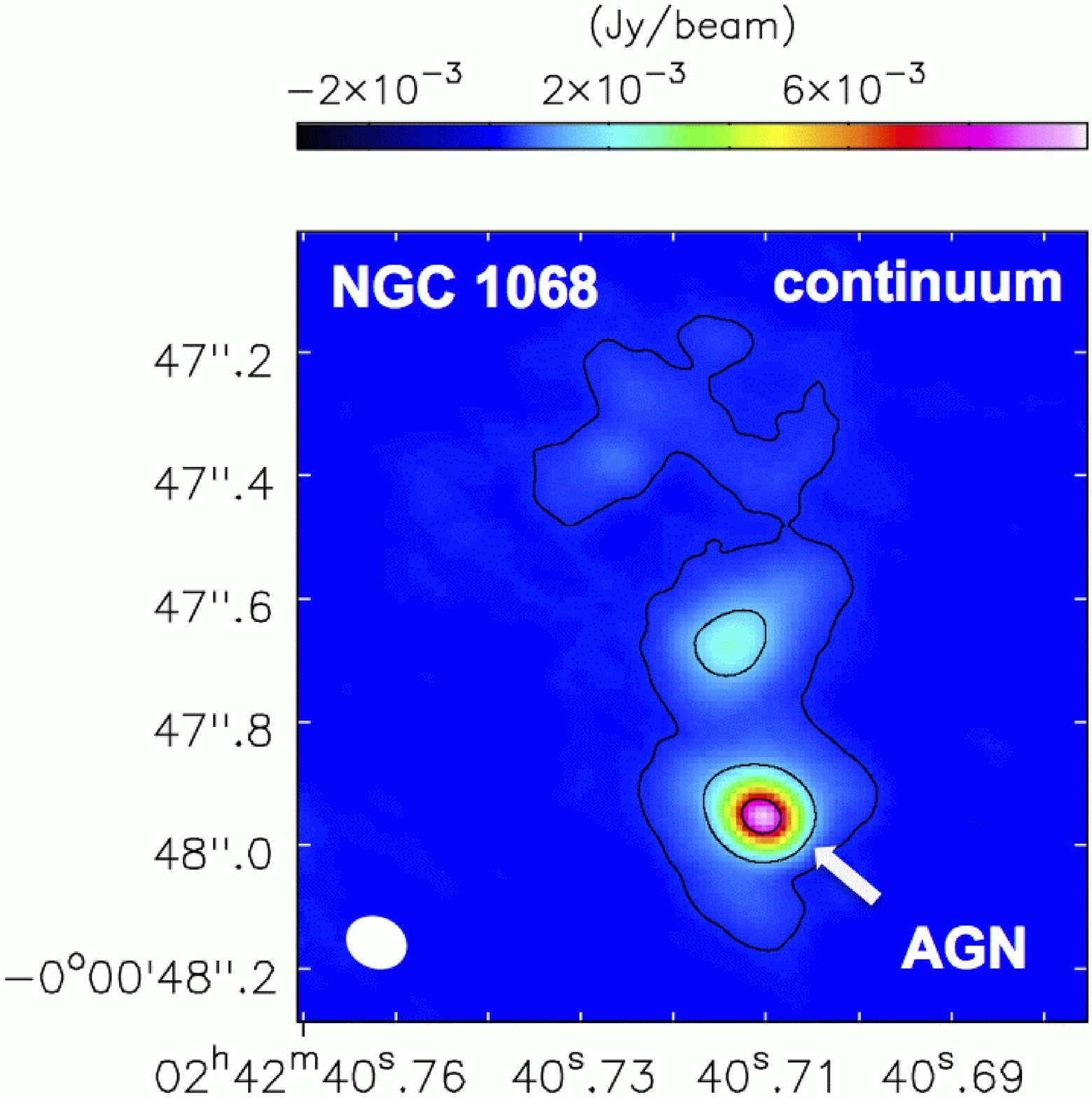} \\ 
\includegraphics[width=7cm]{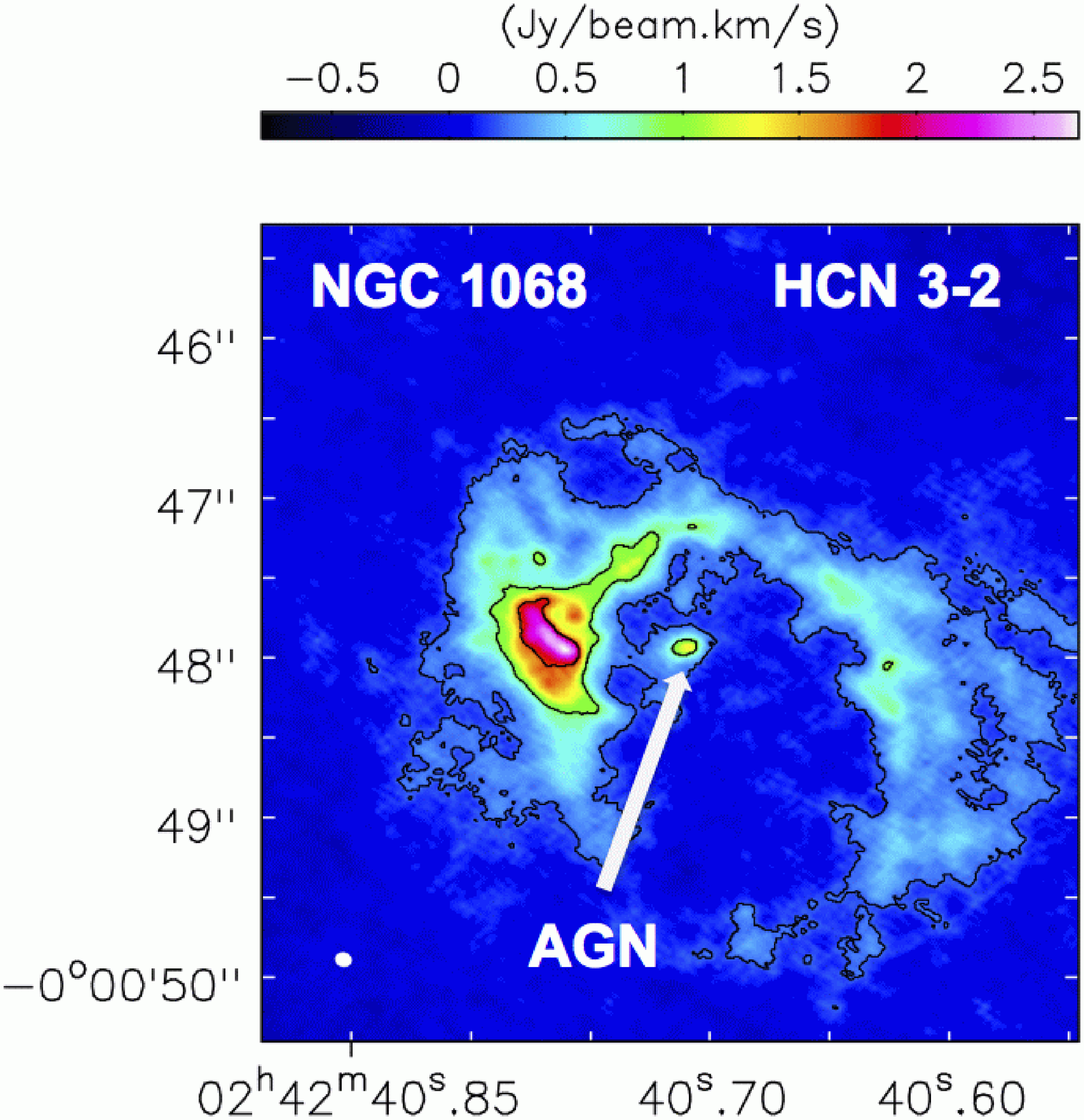} 
\includegraphics[width=7cm]{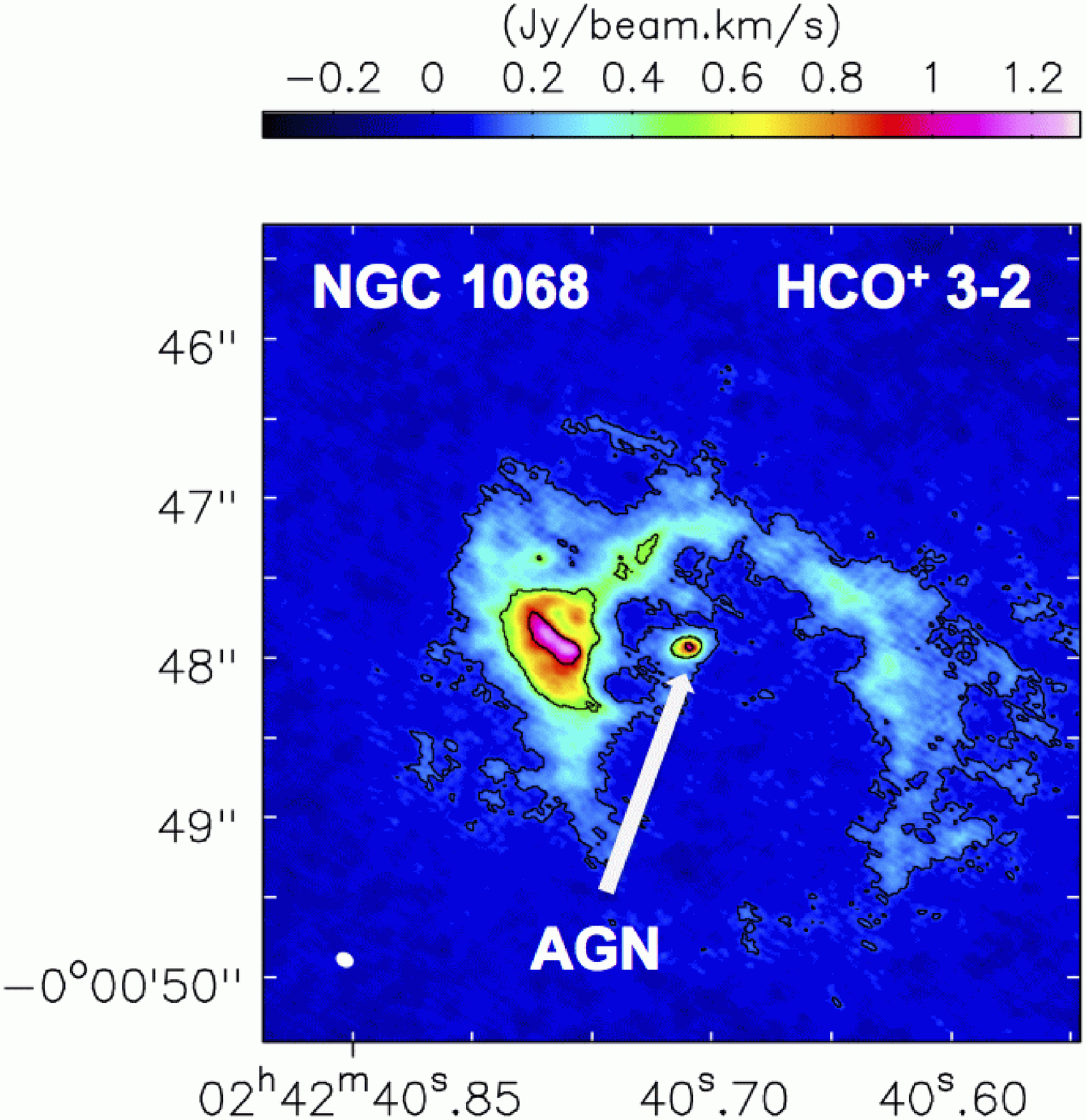} \\ 
\includegraphics[width=7cm]{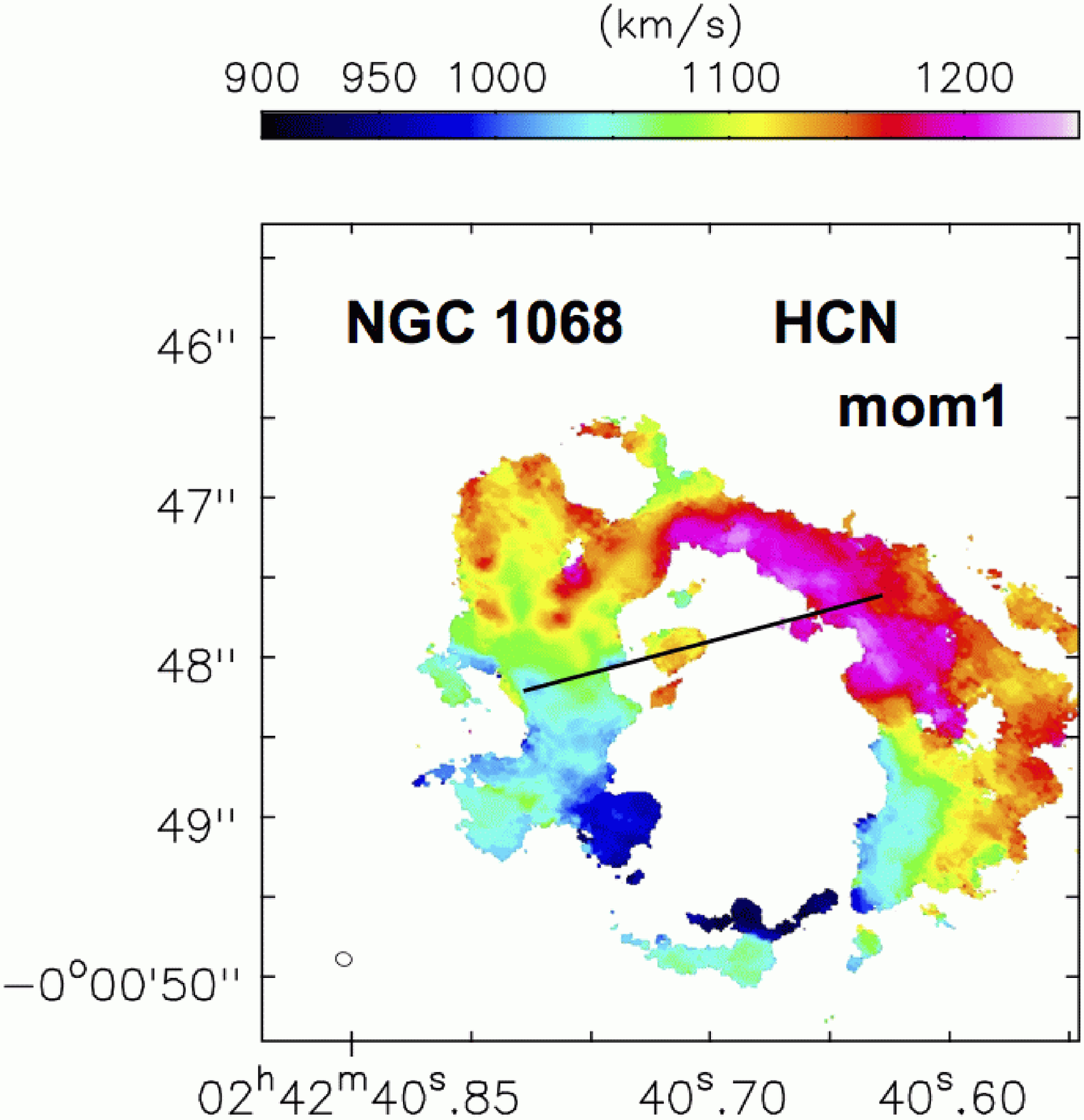} 
\includegraphics[width=7cm]{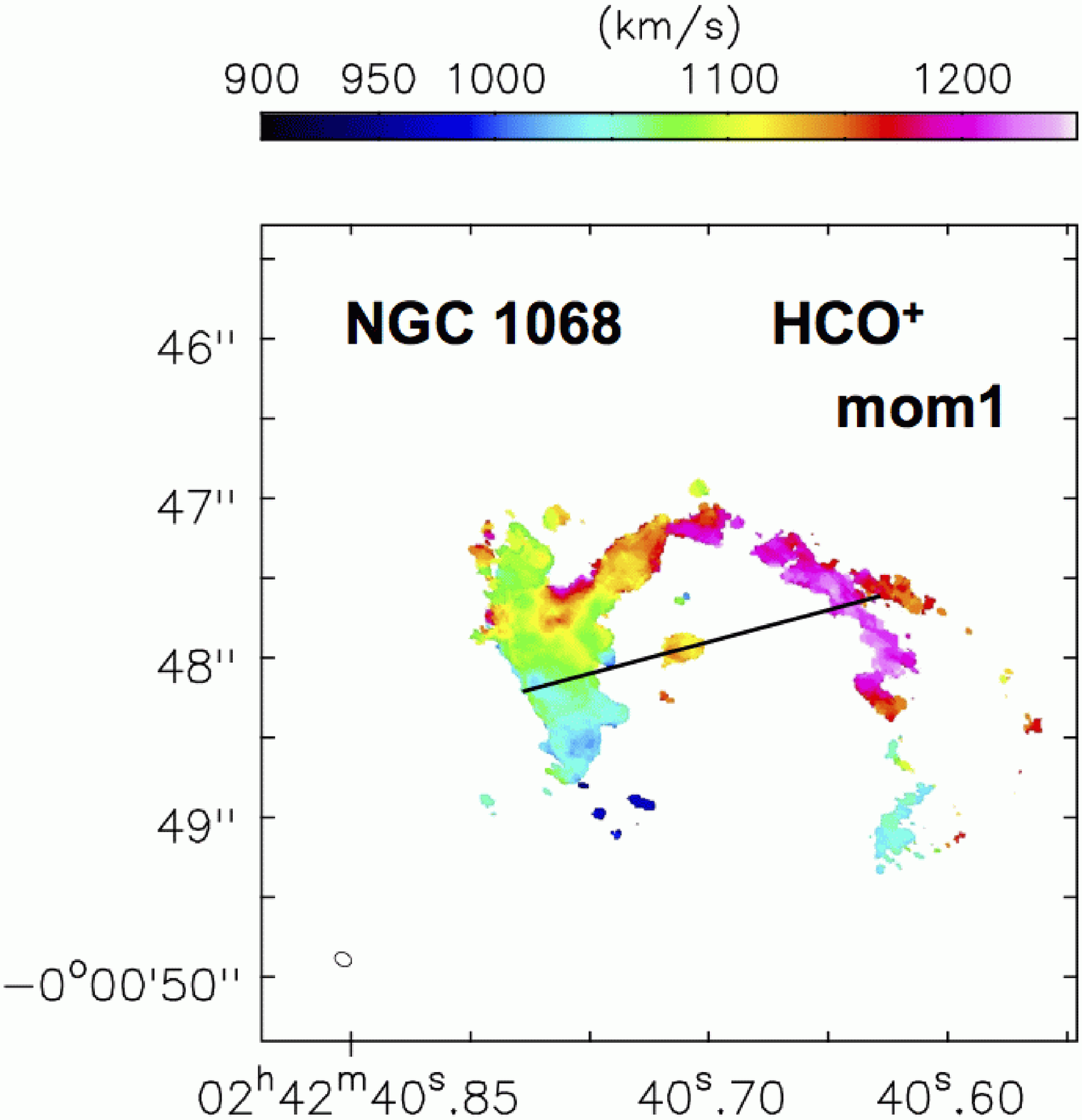} 
\end{center}
\caption{
Continuum emission at $\sim$266 GHz (top), HCN J=3--2 integrated
intensity (moment 0) (middle left), HCO$^{+}$ J=3--2 moment 0 
(middle right), HCN J=3--2 intensity-weighted mean velocity (moment 1)
(bottom left), and HCO$^{+}$ J=3--2 moment 1 maps (bottom right).
All the ALMA Cycle 4 and 2 data were combined. 
The primary beam correction was applied. 
The contours represent 6$\sigma$, 30$\sigma$, and 150$\sigma$ for the
continuum map, and   
3$\sigma$, 12$\sigma$, 24$\sigma$ for the HCN J=3--2 and HCO$^{+}$
J=3--2 moment 0 maps. 
In the moment 1 maps, the solid lines are added at 
PA = 105$^{\circ}$ to indicate the torus direction.
The coordinates are ICRS. 
Optical LSR velocity is used for the moment 1 map.
Since the displayed areas are set to cover significantly detected
emission, the area shown for the continuum data differs from that for
the molecular line data.  
The synthesized beam sizes are 0$\farcs$08--0$\farcs$1 and are
shown as filled or open circles in the lower-left part of each figure.
}
\end{figure}

Figure 2 shows the moment 0, moment 1, and intensity-weighted velocity
dispersion (moment 2) maps of HCN J=3--2 and HCO$^{+}$ J=3--2 for the
nuclear compact emission component (labeled ``AGN'' in Figure 1),
obtained using only the Cycle 4 longest baseline data (a).  
In the moment 0 maps, the emission is elongated almost along the
east--west direction for both HCN J=3--2 and HCO$^{+}$ J=3--2.  
We used the CASA task ``imfit'' function in these maps to estimate the
position angles to be PA = 107$\pm$3$^{\circ}$ and
109$\pm$6$^{\circ}$ (east of north) for the HCN J=3--2 and HCO$^{+}$
J=3--2 emission, respectively. 
The deconvolved emission sizes, which we also estimated using ``imfit'', 
are 186$\pm$17 mas $\times$ 63$\pm$7 mas (13 pc $\times$ 4 pc)
for HCN J=3--2 and 
168$\pm$18 mas $\times$ 71$\pm$12 mas (12 pc $\times$ 5 pc)
for HCO$^{+}$ J=3--2, respectively. 
In the moment 1 maps, we see signatures of rotation patterns 
such that the eastern (western) part of the mass-accreting SMBH is 
redshifted (blueshifted) for both HCN J=3--2 and HCO$^{+}$ J=3--2.
In the moment 2 maps, the velocity dispersion is larger in the western
part than in the eastern part for both HCN J=3--2 and HCO$^{+}$ J=3--2. 
These representative observed properties about morphology and dynamics 
are seen in a similar manner between HCN J=3--2 and HCO$^{+}$ J=3--2,
strongly suggesting that they are real.  

\begin{figure}
\begin{center}
\includegraphics[width=7cm]{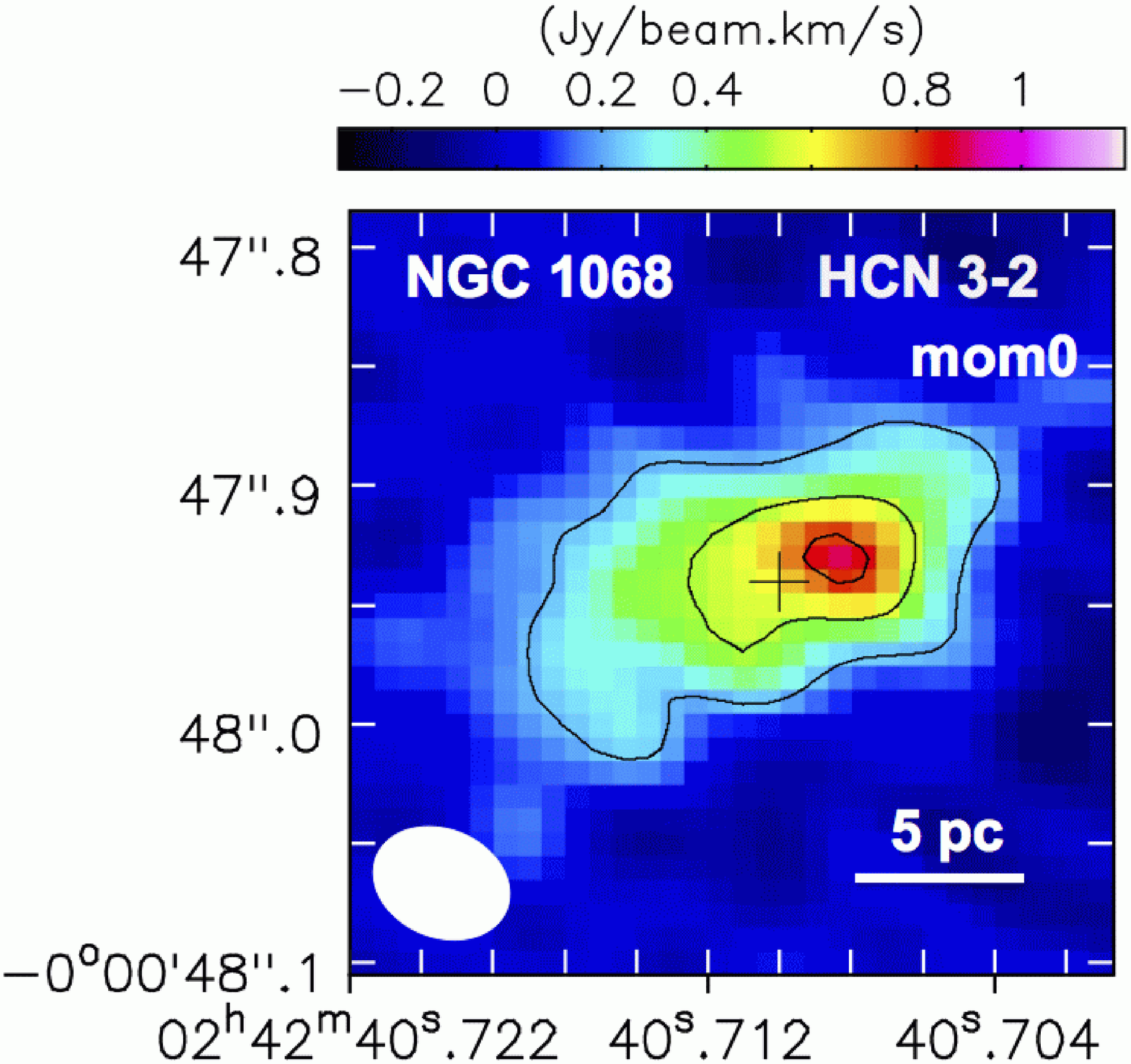} 
\includegraphics[width=7cm]{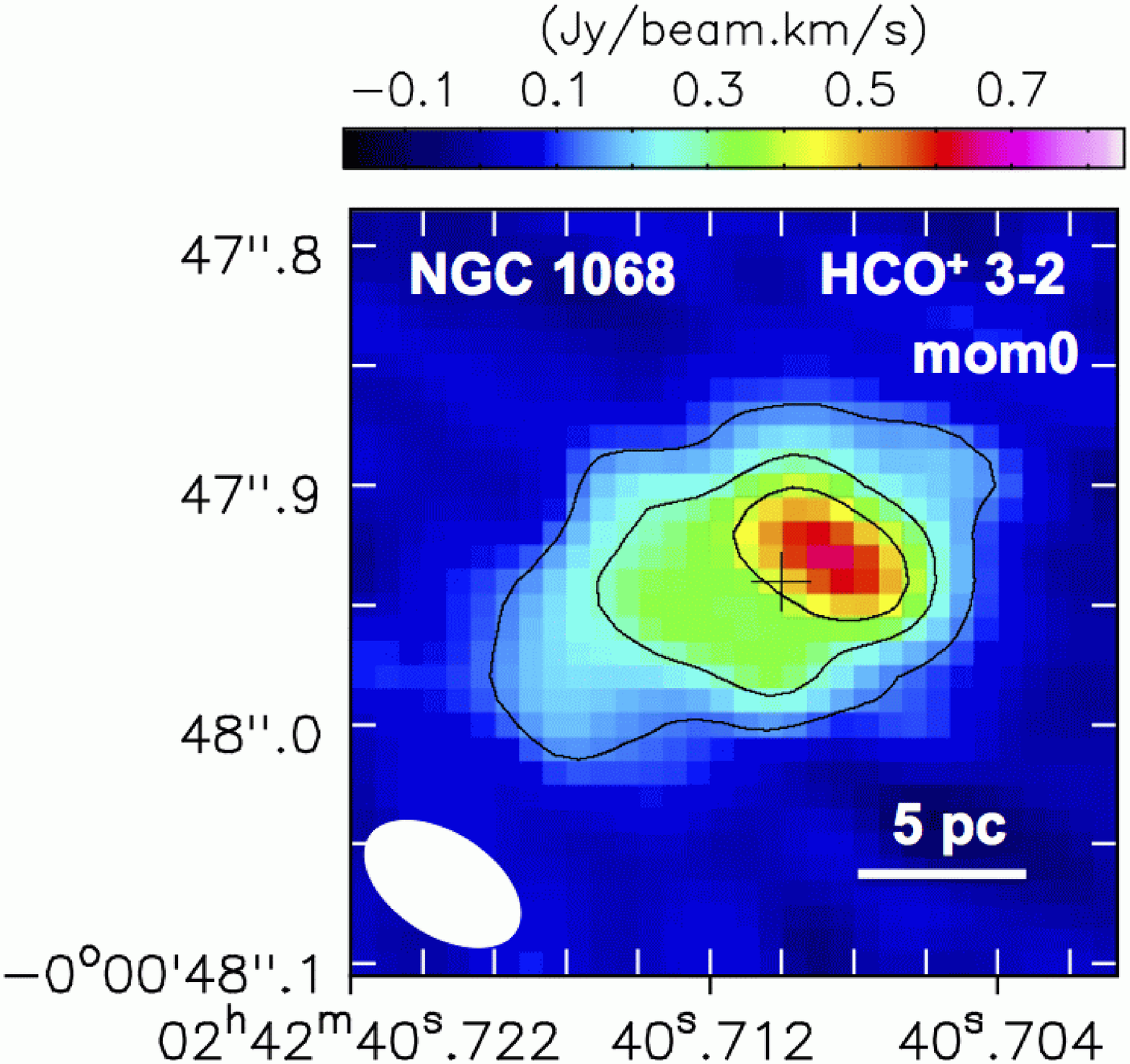} \\
\includegraphics[width=7cm]{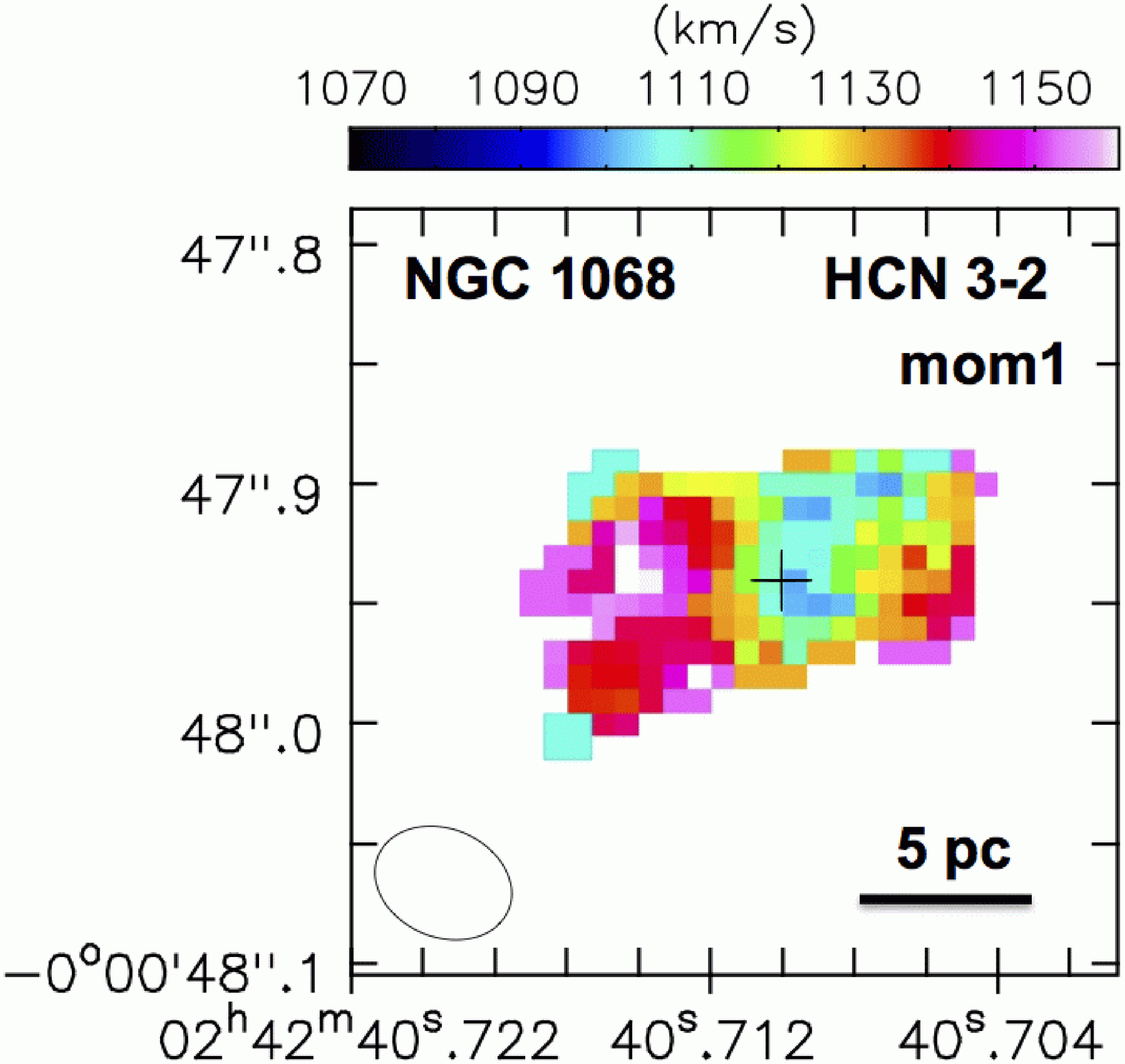} 
\includegraphics[width=7cm]{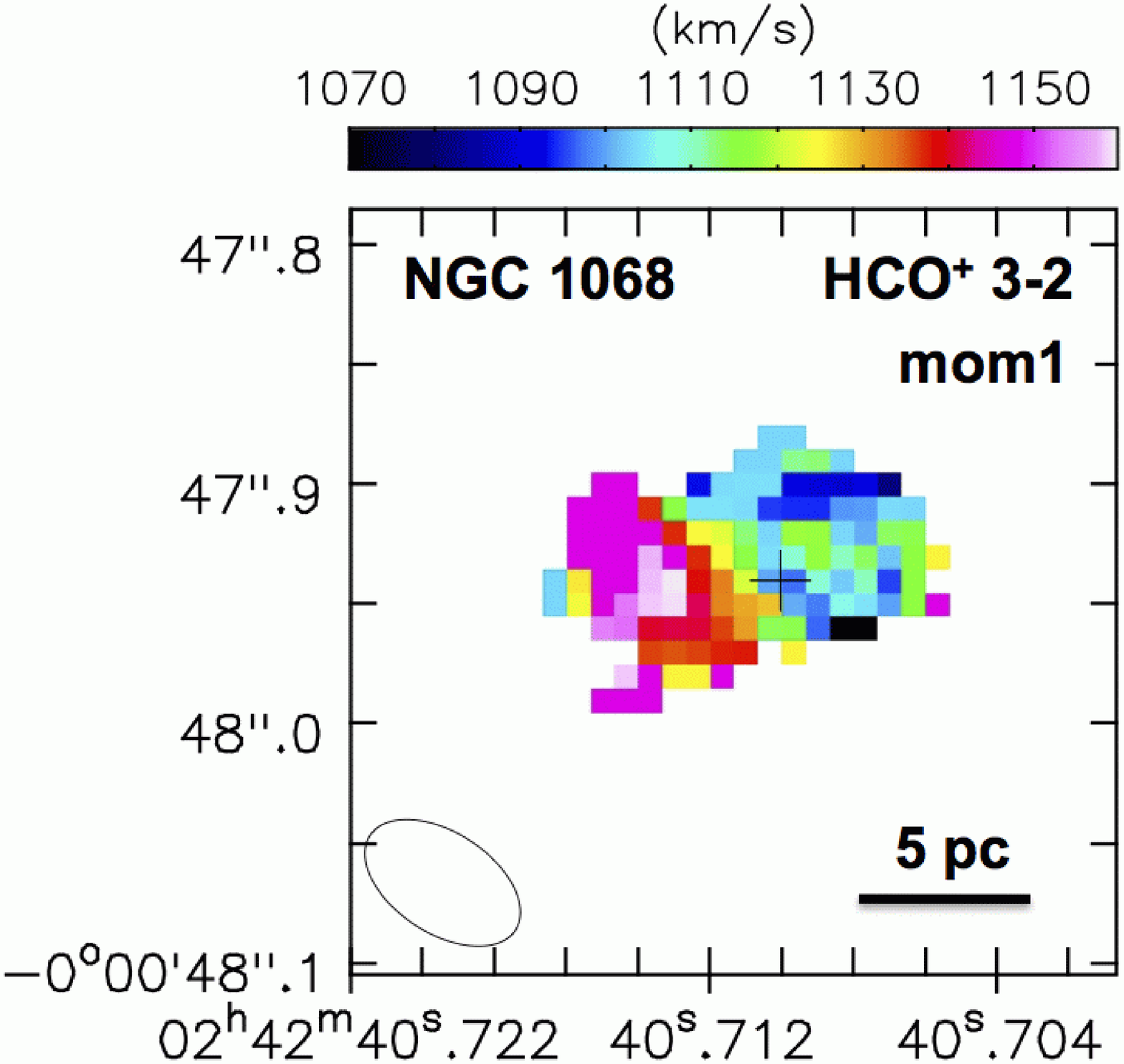} \\
\includegraphics[width=7cm]{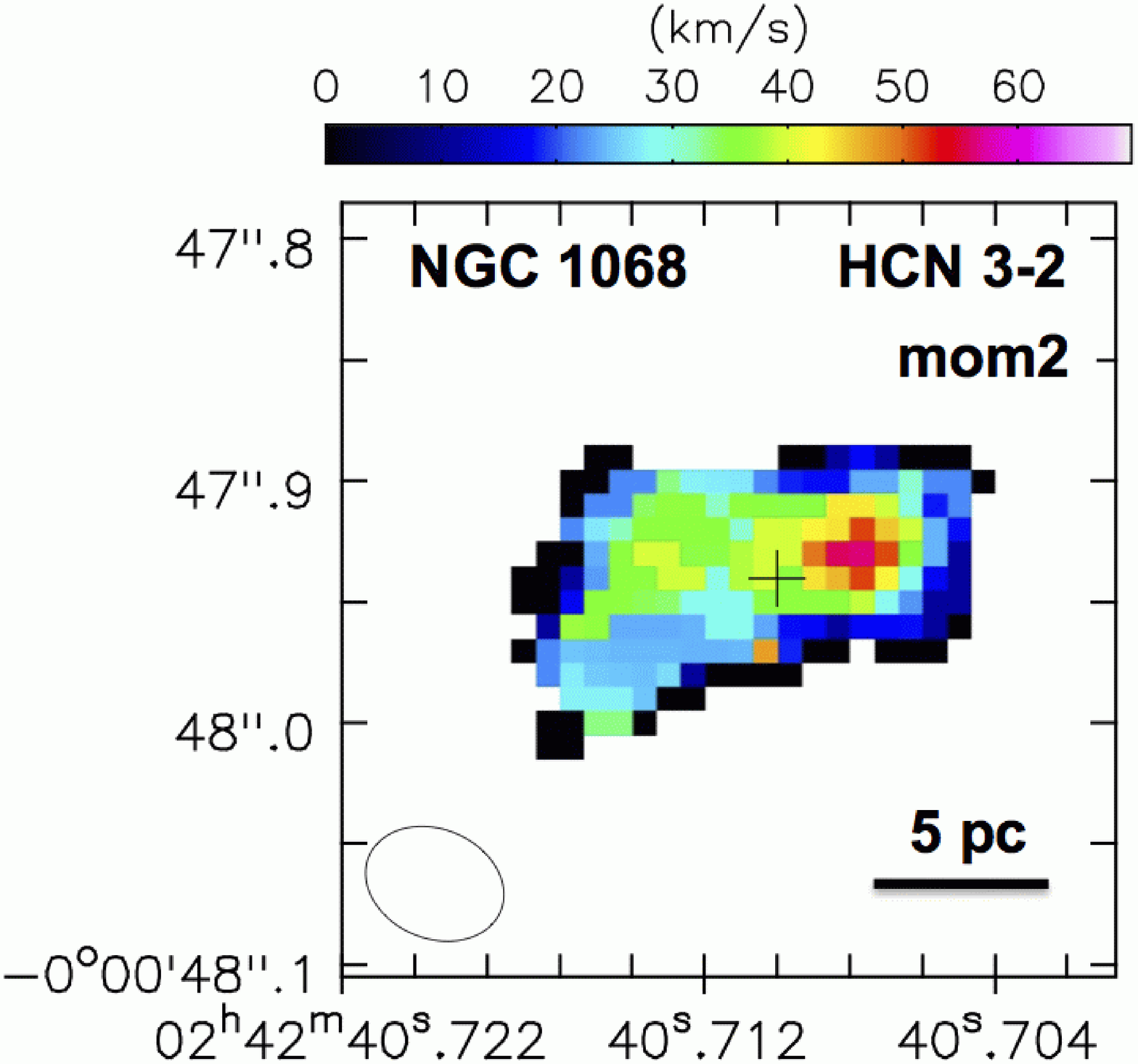} 
\includegraphics[width=7cm]{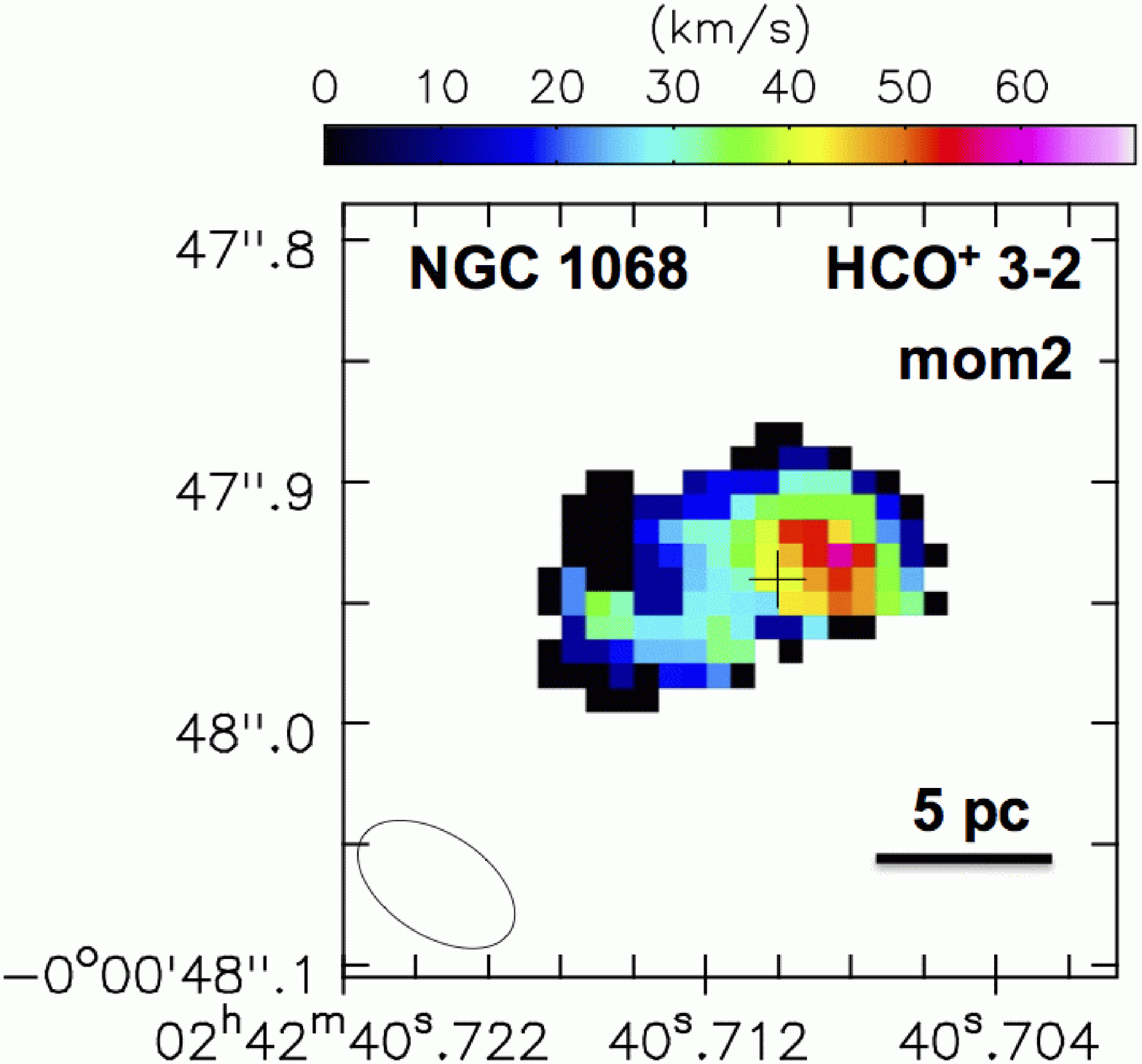} \\
\end{center}
\caption{
\small 
Integrated intensity (moment 0) (Top), intensity-weighted mean
velocity (moment 1) (Middle), and intensity-weighted velocity dispersion 
(moment 2) maps (Bottom) of HCN J=3--2 and HCO$^{+}$ J=3--2 emission
lines for the compact nuclear emission components, created from the data
(a). 
The contours are 3$\sigma$, 6$\sigma$, 9$\sigma$ for the moment 0
maps of HCN J=3--2 and HCO$^{+}$ J=3--2.
For the moment 1 and 2 maps, an appropriate cutoff was applied to prevent
the resulting maps from being dominated by noise.
The coordinates are ICRS. 
The mass-accreting SMBH position at (02$^{h}$42$^{m}$40.710$^{s}$,
$-$00$^{\circ}$00$'$47.94$''$)ICRS  is indicated by black crosses.
The length corresponding to 5 pc is indicated by the solid line. 
The synthesized beam sizes (0$\farcs$04--0$\farcs$07) are shown as
filled or open circles in the lower-left part of each figure.
}
\end{figure}

We investigated the rotational motion in more detail by making spectra
within the beam size at the eastern redshifted region (centered on 
[02$^{h}$42$^{m}$40.714$^{s}$, $-$00$^{\circ}$00$'$47.951$''$]ICRS) and
the western blueshifted region (centered on 
[02$^{h}$42$^{m}$40.708$^{s}$, $-$00$^{\circ}$00$'$47.923$''$]ICRS) 
for HCN J=3--2 (Figure 3a) and HCO$^{+}$ J=3--2 (Figure 3b). 
The peak velocity at the eastern region is slightly redshifted with
respect to that at the western region, as expected from the rotational motion.

We extracted HCN J=3--2 and HCO$^{+}$ J=3--2 spectra within the
0$\farcs$2 (east--west) $\times$ 0$\farcs$15 (north--south) rectangular
region around the mass-accreting SMBH position and applied Gaussian
fits, which are shown in Figure 3(c) and (d).
We obtained peak optical LSR velocity = 1137$\pm$3 (km s$^{-1}$), 
full-width at half-maximum FWHM = 156$\pm$9 (km s$^{-1}$), peak flux
density = 16$\pm$1 (mJy) for HCN J=3--2, and 
peak optical LSR velocity = 1124$\pm$4 (km s$^{-1}$), 
FWHM = 190$\pm$10 (km s$^{-1}$), and peak flux density = 10$\pm$1 (mJy) 
for HCO$^{+}$ J=3--2. 
The Gaussian fit fluxes are 2.6$\pm$0.2 (Jy km s$^{-1}$) and
2.0$\pm$0.1 (Jy km s$^{-1}$) for HCN J=3--2 and HCO$^{+}$ J=3--2,
respectively. 
Some excess emission at the redshifted side of the HCN J=3--2 emission
lines is recognizable (Figure 3c). 

The vibrationally excited HCN v$_{2}$=1f J=3--2 ($\nu_{\rm rest}$ =
267.199 GHz) and HCO$^{+}$ v$_{2}$=1f J=3--2 ($\nu_{\rm rest}$ = 268.689
GHz) emission lines were not clearly detected in the nuclear spectra.

\section{Discussion}


Unlike CO J=6--5 emission, whose morphological and dynamical
directions (roughly east-west and north-south, respectively) are largely
tilted with $>$50$^{\circ}$ \citep{gar16}, our ALMA data show that the
HCN J=3--2 and HCO$^{+}$ J=3--2 emission are aligned approximately along
the east--west direction (PA = 100--110$^{\circ}$), both morphologically
(moment 0 map) and dynamically (moment 1 map).  
This direction is aligned with that of the mid-infrared $\sim$10 $\mu$m
dust continuum emission from the torus
\citep{jaf04,pon06,rab09,bur13,lop14} as well as the VLBA radio
continuum which is thought to trace the inner plasma torus 
\citep{gal04}.     
The inner part of the morphology of the optical [OIII] emission line is
elongated along  the north--south direction \citep{eva91,das06}. 
As AGN's ionizing photons can only escape in directions that are not
blocked by the torus, the [OIII] emission line data also support the
hypothesis that the torus in NGC 1068 is oriented east--west.  
Our high-spatial-resolution ALMA data demonstrate that the HCN J=3--2 and
HCO$^{+}$ J=3--2 emission lines probe the bulk of the rotating
dense molecular gas that spatially coexists with the dust in the torus. 

A simple torus model postulates that the molecular gas and dust are
distributed axisymmetrically around a mass-accreting SMBH.  
However, our ALMA data demonstrate that molecular emission from the
torus is not axisymmetrical in that the western part is brighter and
exhibits larger dispersion than the eastern part for both HCN J=3--2 and
HCO$^{+}$ J=3--2 (Figure 2).
A natural interpretation is that the molecular turbulence is higher in
the western region, which reduces the line opacity, resulting in the
higher observed molecular line flux. 
Molecular gas spatial distribution can be inhomogeneous if conversion
from molecular gas into stars is not uniform spatially. 
Feedback from supernovae in the local nuclear starbursts in the
torus 
\footnote{
Infrared spectroscopy revealed no detectable polycyclic aromatic
hydrocarbon (PAH) emission features (= a good starburst indicator) at
the NGC1068 nucleus \citep{ima97,ima02}. The upper limit of the
starburst far-infrared luminosity is $<$2.7 $\times$ 10$^{43}$ (erg
s$^{-1}$) \citep{ima16} or star-formation rate with $<$1.3 M$_{\odot}$
yr$^{-1}$ \citep{ken98}.  
} 
\citep{wad09} can also alter the spatial distribution and turbulence of
the molecular gas in which case radiation from the central
AGN has a different effect to molecular gas from position to position 
within the torus \citep{wad16}.
The resulting inhomogeneous spatial distribution and dynamics of the
molecular gas, as well as the different AGN radiation effect, may
explain the observed inhomogeneous molecular line emission.  

The HCN J=3--2 luminosity, derived using Equations (1) and (3) of
\citet{sol05}, is 140$\pm$11 [L$_{\odot}$] or
(2.3$\pm$0.2)$\times$10$^{5}$ [K km s$^{-1}$ pc$^{2}$]. 
If the HCN emission is optically thick and thermalized (sub-thermal) at up
to J=3, the luminosity in units of [K km s$^{-1}$ pc$^{2}$] for HCN 
J=1--0 is the same as (larger than) that of HCN J=3--2.
By applying the conversion factor from the HCN J=1--0 luminosity to the dense
molecular mass, M$_{\rm dense}$ = 10 $\times$ HCN J=1--0 luminosity  
[M$_{\odot}$ (K km s$^{-1}$ pc$^{2}$)$^{-1}$] \citep{gao04,kri08}, we
obtain the dense molecular mass within the central
0$\farcs$2$\times$0$\farcs$15 (14 pc $\times$ 10 pc) rectangular region
to be $\sim$2 $\times$ 10$^{6}$ M$_{\odot}$, or possibly even higher. 
Though there may be a large ambiguity in the above conversion
factor, the derived dense molecular mass is substantially smaller than
the SMBH mass of NGC 1068, $\sim$1 $\times$ 10$^{7}$M$_{\odot}$
\citep{gre96,hur02,lod03}, which was estimated from the rotational
motion of a water maser disk much inside the dense molecular
torus that we are now probing.  

From the moment 1 maps in Figure 2 and spectra at the east and west
regions in Figure 3, we derive the rotational velocity to be 
v/sin($i$) $\sim$ 20 (km s$^{-1}$) at distance r $\sim$3 (pc) from
the mass-accreting SMBH, where the inclination angle ($i$) is estimated
to be $i$ = 34--46$^{\circ}$ \citep{gar16}. 
Assuming that this rotation is Keplerian motion governed by the
central SMBH mass, even if we adopt the lowest value of $i$ =
34$^{\circ}$, this yields a value for the enclosed mass inside the
rotating dense molecular disk to be $\sim$9 $\times$ 10$^{5}$M$_{\odot}$,
which is much lower than the above estimated SMBH mass. 
This suggests that the dynamical motion of the dense molecular gas at a
scale of a few pc is far from Keplerian, as argued by some theories
\citep{cha17}. 

At the HCN J=3--2 and HCO$^{+}$ J=3--2 emission lines, 
the eastern (western) part is redshifted (blueshifted) in the torus
(Figure 2), while the eastern (western) dense molecular gas in the
central 0$\farcs$5--2$\farcs$0 region of the host galaxy along the torus 
direction (shown as the solid lines in the bottom plots of 
Figure 1) is blueshifted (redshifted) (Figure 1).
Thus, the torus and molecular gas in the inner part of the host galaxy
are dynamically decoupled, and it seems that some external process is
required. 
\citet{fur16} and \citet{tan17} argued that a minor merger took place in
NGC 1068, and this may be responsible for the dynamical decoupling. 
The dense molecular torus we are probing is likely to have
experienced some disturbance, which might be related to the observed
non-Keplerian motion. 

The amount of molecular gas and dust in the torus in the vicinity close
to the central AGN is expected to decrease due to outflow by 
radiation pressure from the central AGN and nuclear starbursts,
conversion of molecular gas into stars inside the torus,  
a mass inflow into the central SMBH, and/or other forces. 
A mass supply
from the host galaxy is required to maintain the structure of the torus
over the long term.  
We have detected bridging emission from the torus in the eastern
direction to massive molecular clouds in the host galaxy in our new ALMA
data with $>$3$\sigma$, for both HCN J=3--2 and HCO$^{+}$ J=3--2 
(Figure 1).
Even though we cannot rule out the possibility that this emission is just
a superposition of two unrelated emission components along this line of
sight, this might be related to a long-sought-after mass inflow
from a host galaxy to the outer part of an AGN torus. 
Further observations of this emission will be interesting.

\section{Summary}

We conducted 0$\farcs$04--0$\farcs$07 resolution ALMA Cycle 4
observations of the NGC 1068 nucleus at HCN J=3--2 and HCO$^{+}$ J=3--2. 
Thanks to the increase in spatial resolution by a factor of
$\sim$2--3, we have now spatially resolved the compact dense molecular
emission that we had previously detected at the location of the
putative torus at the same molecular lines in our Cycle 2 data. 
We found that (1) the torus-associated dense molecular emission is
elongated along the east--west direction (PA = 100--110$^{\circ}$), 
(2) it is rotating in almost the same direction, but 
with far from pure Keplerian motion governed by the
central SMBH mass, and  
(3) the torus is highly inhomogeneous in that the western
region of the AGN engine is more turbulent and exhibits stronger
molecular line emission than the eastern region does. 
The dense molecular gas in the torus and that in the central 
0$\farcs$5--2$\farcs$0 region of the host galaxy along the torus
direction are counter-rotating, indicating an external process, such as
a minor galaxy merger.

\vspace{0.5cm}

We thank the anonymous referee for his/her valuable comments which  
helped improve the clarity of this manuscript.
We are grateful to Dr. K. Saigo and X. Lu for their supports regarding
ALMA observation preparation and data analysis.  
M.I. is supported by JSPS KAKENHI Grant Number 15K05030.
K.W. is supported by JSPS KAKENHI Grant Number 16H03959.
This paper made use of the following ALMA data:
ADS/JAO.ALMA\#2016.1.00052.S. 
ALMA is a partnership of ESO (representing its member states), NSF (USA) 
and NINS (Japan), together with NRC (Canada), NSC and ASIAA
(Taiwan), and KASI (Republic of Korea), in cooperation with the Republic
of Chile. The Joint ALMA Observatory is operated by ESO, AUI/NRAO, and
NAOJ. 
Data analysis was in part carried out on the open use data analysis
computer system at the Astronomy Data Center, ADC, of the National
Astronomical Observatory of Japan.

\begin{figure}
\begin{center}
\includegraphics[angle=0,scale=.4]{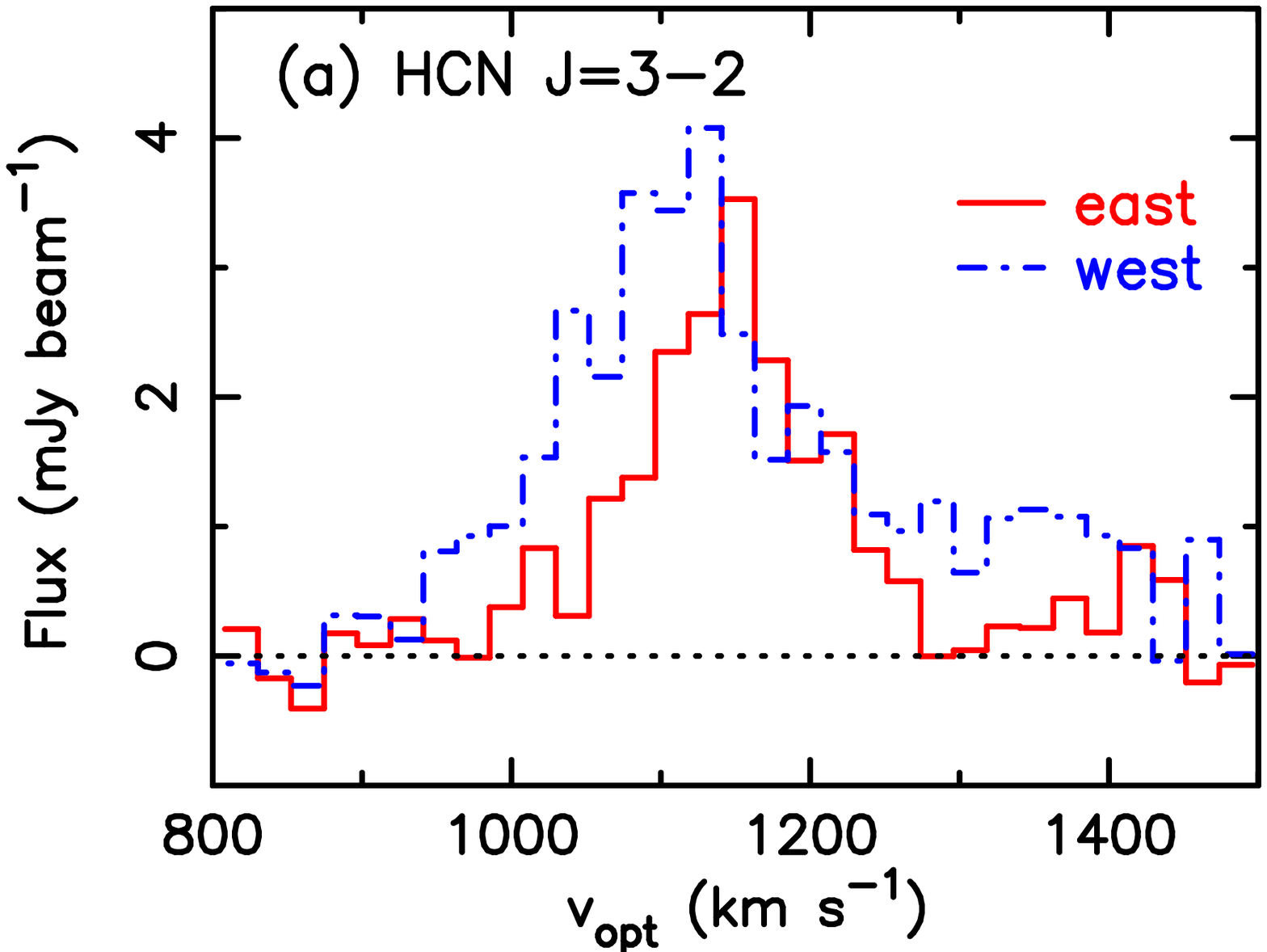} 
\includegraphics[angle=0,scale=.4]{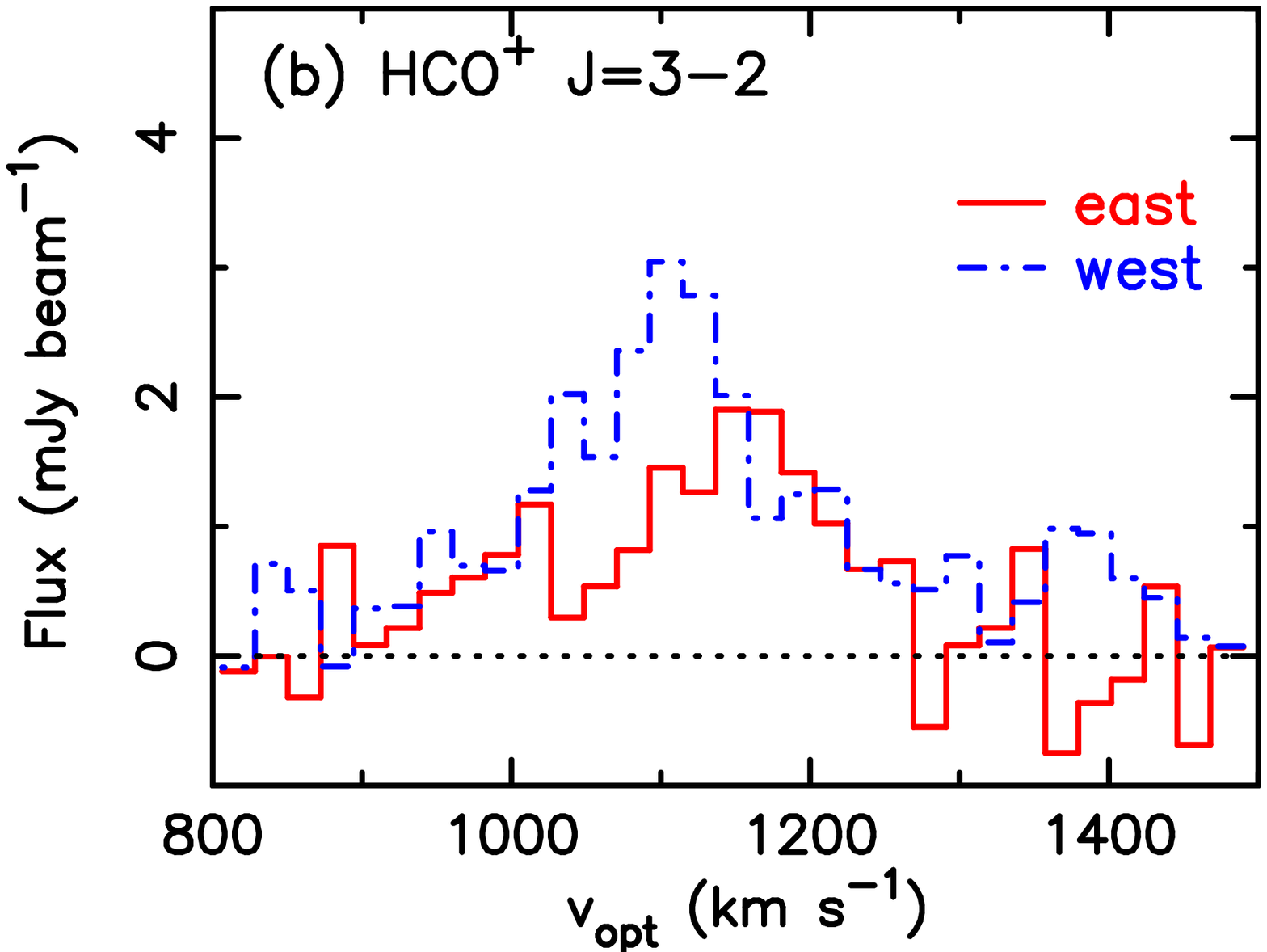} \\  
\includegraphics[angle=0,scale=.4]{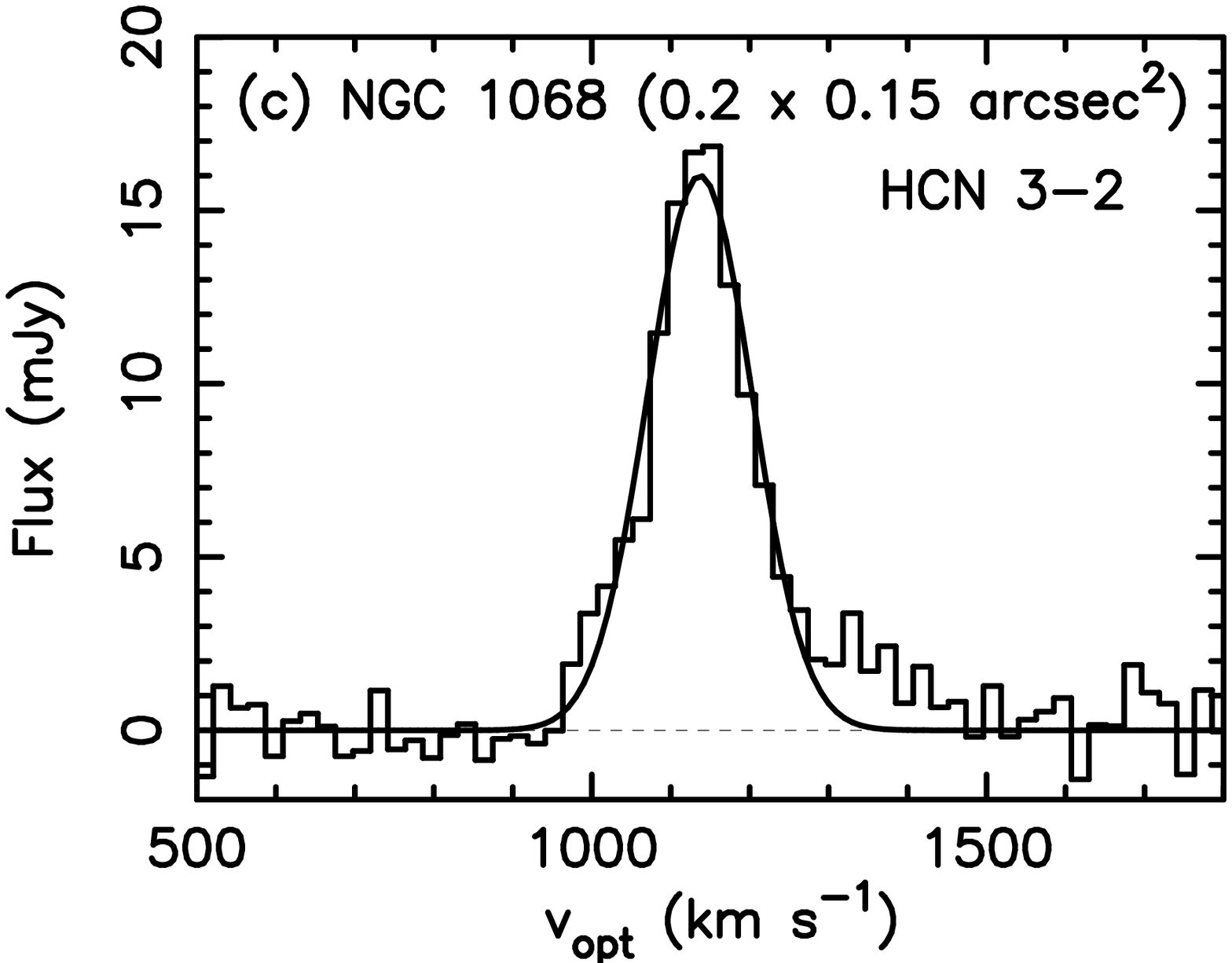} 
\includegraphics[angle=0,scale=.4]{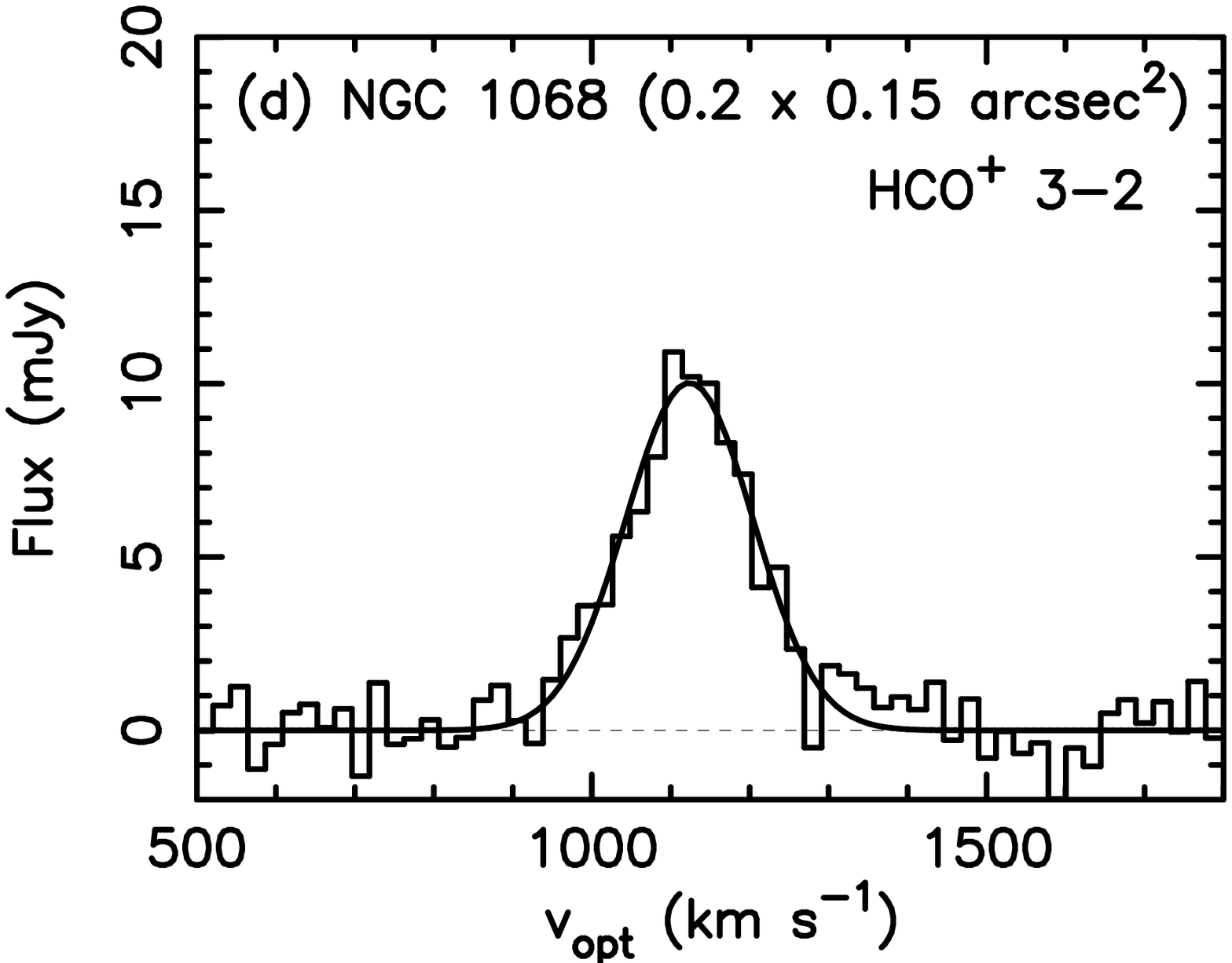} \\  
\end{center}
\caption{
Spectra within the beam size (Figure 2) at the eastern
redshifted (red solid line) and western blueshifted parts (blue
dash-dotted line) of the molecular torus for (a) HCN J=3--2 and (b)
HCO$^{+}$ J=3--2. 
The abscissa is optical LSR velocity, and the ordinate is flux in (mJy
beam$^{-1}$).
Spectra are taken at 
(02$^{h}$42$^{m}$40.714$^{s}$, $-$00$^{\circ}$00$'$47.951$''$)ICRS  
and (02$^{h}$42$^{m}$40.708$^{s}$, $-$00$^{\circ}$00$'$47.923$''$)ICRS
for the eastern and western parts, respectively.
Area-integrated spectra within the central 0$\farcs$2 (east--west)
$\times$ 0$\farcs$15 (north--south) rectangular region, centered on the
SMBH position are shown for (c) HCN J=3--2 and (d) HCO$^{+}$ J=3--2.
The abscissa is optical LSR velocity, and the ordinate is flux in (mJy).
}
\end{figure}


\begin{thebibliography}{}
\bibitem[Alloin et al.(2000)]{alo00}
          Alloin, D., Pantin, E., Lagage, P. O., \& Granato, G. L. 2000,
          A\&A, 363, 926
\bibitem[Antonucci (1993)]{ant93}
         Antonucci, R. R. J. 1993, ARA\&A, 31, 473
\bibitem[Antonucci \& Miller(1985)]{ant85}
         Antonucci, R. R. J. \& Miller, J. S. 1985, ApJ, 297, 621 
\bibitem[Bock et al.(2000)]{boc00}
         Bock, J. J., Neugebauer, G., Matthews, K., 2000, AJ, 120, 2904
\bibitem[Burtscher et al.(2013)]{bur13}
         Burtscher, L., Meisenheimer, K., Tristram, K. R. W., et
         al. 2013, A\&A, 558, A149
\bibitem[Chan \& Krolik(2017)]{cha17}
         Chan, C-H., \& Krolik, J. H. 2017, ApJ, 843, 58
\bibitem[Das et al.(2006)]{das06}
         Das, V., Crenshaw, D. M., Kraemer, S. B., \& Deo, R. P. 2006,
         AJ, 132, 620
\bibitem[Evans et al.(1991)]{eva91}
         Evans, I. N., Ford, H. C., Kinney, A. L. et al., 1991, ApJL,
         369, L27	
\bibitem[Furuya \& Taniguchi(2016)]{fur16}
         Furuya, R., \& Taniguchi, Y., 2016, PASJ, 68, 03
\bibitem[Gallimore et al.(2004)]{gal04} 
         Gallimore, J. F., Baum, S. A., \& O'dea, C. P., 2004, ApJ,
         613, 794  
\bibitem[Gallimore et al.(2016)]{gal16}
         Gallimore, J. F., Elitzur, M., Maiolino, R., et al. 2016,
         ApJL, 829, L7
\bibitem[Gao \& Solomon(2004)]{gao04} 
         Gao, Y., \& Solomon, P. M. 2004, ApJ, 606, 271
\bibitem[Garcia-Burillo et al.(2016)]{gar16}
         Garcia-Burillo, S., Combes, F., Ramos Almeida, C., et al. 2016,
         ApJL, 823, L12
\bibitem[Greenhill et al.(1996)]{gre96}
         Greenhill, L. J., Gwinn, C. R., Antonucci, R., \& Barvainis,
         R. 1996, ApJ, 472, L21
\bibitem[Honig et al.(2017)]{hon17}
         Honig, S. F., \& Kishimoto, M. 2017, ApJL, 838, 20
\bibitem[Hure(2002)]{hur02}
         Hure, J. -M. 2002, A\&A, 395, L21
\bibitem[Imanishi(2002)]{ima02} 
         Imanishi, M. 2002, ApJ, 569, 44
\bibitem[Imanishi et al.(2016)]{ima16}
          Imanishi, M., Nakanishi, K., \& Izumi, T. 2016, ApJL, 822, L10   
\bibitem[Imanishi et al.(1997)]{ima97}
         Imanishi, M., Terada, H., Sugiyama, K., et al. 1997, PASJ, 49,
         69
\bibitem[Jaffe et al.(2004)]{jaf04} 
          Jaffe, W., Meisenheimer, K., Rottgering, H. J. A., et
          al. 2004, Nature, 429, 47
\bibitem[Kennicutt(1998)]{ken98}
         Kennicutt, Jr. R. C. 1998, ARA\&A, 36, 189
\bibitem[Krips et al.(2008)]{kri08}
         Krips, M., Neri, R., Garcia-Burillo, S., Martin, S., Combes,
         F., Gracia-Carpio, J., \& Eckart, A. 2008, ApJ, 677, 262 
\bibitem[Lodato \& Bertin(2003)]{lod03}
         Lodato, G., \& Bertin, G. 2003, A\&A, 398, 517
\bibitem[Lopez-Gonzaga et al.(2014)]{lop14}
         Lopez-Gonzaga, N., Jaffe, W., Burtscher, L., Tristram,
         K. R. W., \& Meisenheimer, K. 2014, A\&A, 565, A71
\bibitem[Poncelet et al.(2006)]{pon06}
         Poncelet, A., Perin, G., \& Sol, H. 2006, A\&A, 450, 483
\bibitem[Raban et al.(2009)]{rab09} 
         Raban, D., Jaffe, W., Rottgering, H., et al. 2009, MNRAS, 394,
         1325 
\bibitem[Schartmann et al.(2008)]{sch08}
         Schartmann, M., Meisenheimer, K., Camenzind, M., et al. 2008,
         A\&A, 482, 67 
\bibitem[Solomon \& Vanden Bout(2005)]{sol05}
         Solomon, P. M., \& Vanden Bout, P. A. 2005, ARA\&A, 43, 677
\bibitem[Tanaka et al.(2017)]{tan17}
          Tanaka, I., Yagi, M., \& Taniguchi, Y., 2017, PASJ, 69, 90
\bibitem[Tomono et al.(2001)]{tom01}
          Tomono, D., Doi, Y., Usuda, T., \& Nishimura, T. 2001, ApJ,
          557, 637
\bibitem[Tristram et al.(2014)]{tri14}
          Tristram, K. R. W., Burtscher, L., Jaffe, W. et al., 2014,
          A\&A, 563, 82
\bibitem[Wada et al.(2009)]{wad09}
         Wada, K., Papadopoulos, P. P., \& Spaans, M. 2009, ApJ, 702, 63
\bibitem[Wada et al.(2016)]{wad16}
          Wada, K., Schartmann, M., \& Meijerink, R., 2016, ApJL, 828,
          L19
\end{thebibliography}
\end{document}